\documentclass[a4paper,twocolumn]{article}

\usepackage{cite}

\usepackage{amsmath,amssymb,amsfonts}
\usepackage{algorithmic}
\usepackage{graphicx,color}
\usepackage{textcomp,bm}
\usepackage{xcolor}
\usepackage[applemac,utf8]{inputenc}
\usepackage{svg}
\usepackage[hidelinks]{hyperref}
\usepackage{algorithm,algorithmic}
\usepackage{geometry}
\usepackage[font=small,skip=10pt]{caption}

\usepackage{glossaries}

\def\BibTeX{{\rm B\kern-.05em{\sc i\kern-.025em b}\kern-.08em
    T\kern-.1667em\lower.7ex\hbox{E}\kern-.125emX}}
\AtBeginDocument{\definecolor{tmlcncolor}{cmyk}{0.93,0.59,0.15,0.02}\definecolor{NavyBlue}{RGB}{0,86,125}}
\usepackage{titlesec}

\title{Chirp-Based Ambiguity Function Analysis for Discrete Arrays in the Spherical Wavefront Regime}
\title{\textsc{Chirp-Based Aliasing Analysis of Arrays in the Spherical Wavefront Regime }}

\author{G. Monnoyer, L. Defraigne, B. Sambon, J. Louveaux, L. Vandendorpe \vspace{8pt} \\
\textit{ICTEAM, UCLouvain, Louvain-la-Neuve (Belgium)}, \\
\textit{emails~: firstname.lastname@uclouvain.be}}

\date{\vspace{5pt}}

\begin{document}
\newcommand{\cc}{\centering}
\newcommand{\bs}{\boldsymbol}
\newcommand{\fs}{\mathsf}
\newcommand{\bb}{\mathbb}
\newcommand{\cl}{\mathcal}
\newcommand{\bcl}[1]{\boldsymbol{\mathcal{#1}}}

\newcommand{\ts}{\textstyle}
\newcommand{\ie}{\emph{i.e.}, }
\newcommand{\eg}{\emph{e.g.}, }
\newcommand{\etal}{\emph{et al.}}
\newcommand{\st}{\ensuremath{\mathrm{s.t.}}\xspace}

\newcommand{\GM}[1]{
{\color{blue} [ {\textbf GM : } {\small #1} ]}
}

\newcommand{\BS}[1]{
{\color{orange} [ {\textbf BS : } {\small #1} ]}
}

\newcommand{\LD}[1]{
{\color{purple} [ {\textbf LD : } {\small #1} ]}
}

\newcommand{\LV}[1]{
{\color{red} [ {\textbf LV : } {\small #1} ]}
}
\newcommand{\unitvec}{\hat{\bs e}}
\newcommand{\anglecasual}{\phi}
\newcommand{\indicator}{\eta}
\newcommand{\bessel}{\mathrm{J}}
\newcommand{\sinc}{\mathrm{sinc}}

\newcommand{\mtest}[1]{{\tilde{#1}}}
\newcommand{\mapprox}[1]{{\hat{#1}}}
\newcommand{\ca}{{\mathrm{ca}}}
\newcommand{\ula}{{\mathrm{ula}}}

\newcommand{\antloc}{{\bs x}}
\newcommand{\antlock}{\antloc^*}
\newcommand{\antradius}{R}
\newcommand{\antangle}{\theta}
\newcommand{\sourceloc}{{\bs x_\mathrm{s}}}
\newcommand{\sourceradius}{R_\mathrm{s}}
\newcommand{\sourceangle}{\theta_\mathrm{s}}
\newcommand{\testedloc}{{\mtest{\bs x}_\mathrm{s}}}
\newcommand{\testedradius}{\mtest R_\mathrm{s}}
\newcommand{\testedangle}{\mtest \theta_\mathrm{s}}
\newcommand{\diffradius}{R_{\mathrm{s} \mtest{\mathrm{s}}}}
\newcommand{\diffradiusmax}{R_{\mathrm{s} \mtest{\mathrm{s}}, \mathrm{max}}}
\newcommand{\diffangle}{\theta_{\mathrm{s} \mtest{\mathrm{s}}}}
\newcommand{\antdomain}{\mathcal X_\mathrm{A}}
\newcommand{\antradiusdomain}{\mathcal{R}_\mathrm{A}}
\newcommand{\antangledomain}{\Theta_\mathrm{A}}
\newcommand{\sourcedomain}{\mathcal X_\mathrm{S}}

\newcommand{\antk}{{\bs k}}
\newcommand{\antkradius}{\rho}
\newcommand{\antkangle}{\phi}
\newcommand{\sourcek}{{k_\mathrm{s}}}
\newcommand{\sourcewavelen}{\lambda_\mathrm{s}}
\newcommand{\sourcephase}{\phi_\mathrm{s}}
\newcommand{\spatialband}{K}

\newcommand{\sigtx}{s}
\newcommand{\sigrx}{s_\mathrm{R}}
\newcommand{\sig}{z}
\newcommand{\spectrum}{\mathcal Z}

\newcommand{\sigprod}{g}
\newcommand{\sigprodphase}{\phi}
\newcommand{\sigprodamp}{\alpha}
\newcommand{\spectrumprod}{G}

\newcommand{\antspacing}{\delta}
\newcommand{\caradius}{R_\ca}
\newcommand{\aperture}{\psi}
\newcommand{\visualaperture}{\Omega_\ca}
\newcommand{\lenula}{L}

\newcommand{\ambifun}{A}

\newcommand{\stset}{\mathcal S}
\newcommand{\weight}{w}
\newcommand{\seppar}{ ; \,}
\newcommand{\paramvar}{\tau}
\newcommand{\paramfreq}{k_\tau}
\newcommand{\paramdomain}{\mathcal T}

\newacronym{af}{AF}{Ambiguity Function}

\newacronym{ft}{FT}{Fourier transform}

\newacronym{aae}{AAE}{Antenna Array Element}

\newacronym{nf}{NF}{Near Field}
\newacronym{ff}{FF}{Far Field}
\newacronym{swr}{SWR}{Spherical Wavefront Regime}
\newacronym{sdma}{SDMA}{Space Division Multiple Access}

\newacronym{ula}{ULA}{Uniform Linear Array}
\newacronym{ca}{CA}{Circular Array}
\newacronym{mimo}{MIMO}{Multiple-Inputs Multiple-Outputs}
\newacronym{mas}{MAS}{Multiple Antenna System}

\titleformat{\section}
  {\normalfont\normalsize\scshape}{\thesection.}{1em}{}
\titleformat{\subsection}
  {\normalfont\normalsize\it\sffamily}{\thesubsection.}{1em}{}

\maketitle

\begin{abstract} 
\textbf{ \small{
In antenna arrays, wave propagation modeling based on Euclidean principles is typically represented by steering vectors or signals. 
This paper provides a new, chirp-based, interpretation of steering vectors in the \gls{swr}, establishing a relationship between the spatial spectrum of the received (resp. transmitted) signal and the geometry of the array and the source (resp. target).
Leveraging the well-known sampling theorem, we analyze aliasing effects arising from spatial sampling with a finite number of antennas and understand how these effects degrade the \gls{af}. 
Our framework provides geometric insight into these degradations, offering a deeper understanding of the non-space-invariant aliasing mechanisms in the \gls{swr}. 
The proposed approach is formulated for general antenna arrays and then instantiated to \acrlong{ca} and to \acrlong{ula} structures operating in \acrlong{nf} conditions.}}
\end{abstract}

\glsresetall
\section{Introduction}

\glspl{mas} are essential for both communication and positioning applications.
\gls{mas} are anticipated to support future multi-functional wireless networks~\cite{chen_6g_2024}. 
Two key advancements distinguish modern \glspl{mas} from legacy systems: the use of higher frequency bands, including the THz range, and the expansion of antenna array sizes~\cite{liu_near_field_2023}.

Before these advancements, operating conditions defined by the Fraunhofer distance permitted a \gls{ff} wave modeling, approximating the true spherical wavefront as planar. In \gls{ff} conditions, the angle of arrival or departure, associated with the so-called directional cosine, is the sole parameter controllable by the antenna array. Space Division Multiple Access (SDMA)
exploits this parameter to separate users. 

The expansion of antenna array sizes necessitates accounting for the true spherical nature of wavefronts~\cite{zhang_6g_2023}. 
This implies that a single array captures or transmits steering signals containing both range and angular information. Arrays operating in the \gls{swr} are consequently capable of positioning or beam-pointing~\cite{demir_foundations_2021}, and of separating users in both the range and the angle domains~\cite{wu_multiple_2023}. 
While the \gls{ff} directional cosine is associated with a single spatial frequency (or wave number), it has been noticed for instance in~\cite{kosasih_spatial_2025} that \glspl{ula} operating in \gls{nf} conditions capture a wider spatial spectrum influenced by both the source's angle and range. 

This broader spectrum arises from the combination of local sub-arrays, each associated with one directional cosine, hence motivating the interpretation of the steering vector as defining a \emph{spatial chirp}.
In the time domain, chirps relate to the concept of instantaneous frequency. 
In the \gls{swr}, the spatial chirp approach introduces local wave vectors (local planar waves) instead of ``global ones" as associated with the Fourier transform~\cite{chassande_stationary_1998}. 
The observation of a chirp-like structure in \gls{nf} communication modeled by the familiar parabolic or Fresnel wavefront approximation has been reported in \cite{swindlehurst_passive_1988, qiu_doa_2018, jian_fractional_2024}. 
To the best of our knowledge, a chirp-based analysis has not yet been exploited for general wavefronts to understand the impact of the array and source geometry on the achievable performance.

This paper therefore proposes to use the chirp structure of steering vectors associated with \gls{swr} arrays to study the achievable \gls{af} --or interference pattern\cite{vandendorpe_positioning_2025}.
We introduce a tractable methodology to capture the complex interplay between the \gls{swr} array geometry and aliasing patterns in the \gls{af}.
We expect this to help in designing antenna arrays that match the desired performance in the \gls{swr}.

As a result, this paper  
\begin{itemize}
    \item develops a chirp-based representation of steering vectors;
    \item leverages this chirp aspect and the sampling theorem to understand the geometry of the aliasing mechanisms appearing in the \gls{af}, given a finite number of antennas; 
    \item formalizes its findings (in section~\ref{sec:chirp-methodology}) in a framework that generalizes standard \gls{ff} rules to the \gls{nf} context;
    \item and applies it to canonical \gls{ca} and \gls{ula} structures in the \gls{swr} (in sections~\ref{sec:ca} and \ref{sec:ula}).
\end{itemize}

\section{System model and wavefront} 

We consider a static uplink scenario in which a source located in $\sourceloc$ transmits a signal $\sigtx(t)$ captured by a space-continuous antenna array contained in domain $\antdomain$.
We assume that $\sigtx(t)$ occupies a band centered around a carrier frequency associated with a wave number $\sourcek = 2\pi/\sourcewavelen$, given the carrier wavelength $\sourcewavelen$.
In the current paper, we assume a narrowband signal, enabling us to focus on the wave propagation aspects.
Following physical optics principles, the propagation coefficient between the source and the \gls{aae} located in $\antloc\in\antdomain$ is expressed as
\begin{equation}
\label{eq:signal-model}
\sig(\antloc ; \sourceloc)=
\frac{ \exp(-j\sourcek \| \antloc -\sourceloc \| )}
{\| \antloc -\sourceloc \| }.
\end{equation}

In words, $\sig(\antloc ; \sourceloc)$ provides the phase and attenuation received at the antenna located in position $\antloc$, parameterized by the given source location $\sourceloc$.
The \gls{swr}, or \gls{nf} context, captured by \eqref{eq:signal-model}, results in new rules for the aliasing and resolution
inherent to the acquisition process. 

\section{Chirp-based ambiguity function analysis}
\label{sec:chirp-methodology}
 
The \gls{af} is an important tool to evaluate the estimation performance achievable in view of the structural dependence of the observations on the parameter(s) to be estimated (in our case: $\sourceloc$). 
Motivated by a Maximum Likelihood Estimation Approach in the presence of Additive White Gaussian Noise, the \gls{af} provides the noise-free output of a filter $\antloc$-matched to the \emph{tentative} position $\testedloc$ while the \emph{true} source's position is $\sourceloc$. 
By definition, and assuming a narrow-band signal, the \gls{af} associated with our context is given by
\begin{equation}
    \label{eq:ambifun}
    \ambifun(\testedloc, \sourceloc) 
    :=
    \int_{\antdomain}
    \sig(\antloc \seppar \sourceloc)
    \sig^*(\antloc \seppar \testedloc)
    \mathrm{d} \antloc.
\end{equation}

Finite antenna arrays yield an \gls{af} expressed as a summation instead of the integral.
Too coarse sampling can cause aliasing, such as the well-known angular repetition appearing in the \gls{ff} of \glspl{ula} when the antenna spacing is larger than the half-wavelength.
With this paper, we provide a framework to generalize this effect to the \gls{nf} context. 
Using a chirp-based analysis, we obtain tractable closed-form expressions for the aliasing rules in \gls{nf}.

\subsection{Aliasing in the ambiguity function}
We restrict the study of this paper to cases where $\antdomain$ draws a curve in $\mathbb R^2$, described using a parametric variable $\paramvar$ as
\begin{equation}
    \label{eq:domain-as-line}
    \antdomain = \{ \antloc(\paramvar) : \paramvar \in \paramdomain \},
\end{equation}
for a given compact interval $\paramdomain \in \mathbb R$.
We leave the study of multi-dimensional arrays to future contributions. 

We are interested in characterizing the chirp structure of the function 
\begin{align}
    \label{eq:def-sigprod}
    \sigprod(\paramvar \seppar \testedloc, \sourceloc) 
    &:= \sig\big(\antloc(\paramvar) \seppar \sourceloc\big) 
    \, \sig^*\big(\antloc(\paramvar) \seppar \testedloc\big)
    \left\| \dot \antloc (\paramvar) \right\|
\end{align}
where $\dot \antloc(\paramvar)$ denotes the derivative of $\antloc(\paramvar)$ with respect to $\paramvar$. 
This will enable us to approximate the support of its Fourier transform with respect to the variable $\paramvar$, which is given by
\begin{equation}
    \label{eq:sigprod-ft}
    \spectrumprod(\paramfreq \seppar \testedloc, \sourceloc) 
    :=
    \int_{\paramdomain}
    \sigprod(\paramvar \seppar \testedloc, \sourceloc)
    \exp(-j \paramfreq \paramvar) 
    \mathrm{d} \paramvar.
\end{equation}

The key observation for our study is that the \gls{af} in~(\ref{eq:ambifun}) is exactly the above spectrum $\spectrumprod$ evaluated in $\paramfreq = 0$.
Discretizing $\antdomain$ in $N$ points by uniformly sampling $\paramdomain$ with step $\antspacing$ folds the spectrum $\spectrumprod$ with a period $2\pi\antspacing^{-1} = 2\pi\frac{N}{|\paramdomain|}$.
The \gls{af} will only be modified by the discretization if the folding affects the spectrum in $\paramfreq = 0$.
Since we can afford folding anywhere else on $\spectrumprod$, the \emph{sufficient} condition that ensures no such specific spectral zero-folding is softer than the Shannon-Nyquist sampling theorem by a factor of 2. 

More precisely, given a specific pair $\sourceloc$, $\testedloc$, and defining the spatial band limit\footnote{In practice, we commonly use a small tolerance $\epsilon>0$ instead of 0 in the definition \eqref{eq:def-band-limit} to avoid $\spatialband(\testedloc, \sourceloc) = \infty$} as
\begin{equation}
    \label{eq:def-band-limit}
    \spatialband(\testedloc, \sourceloc) = \max\{|\paramfreq|: |\spectrumprod(\paramfreq)| > 0\},
\end{equation}
it is guaranteed that no zero-folding happens when 
\begin{equation}
    \label{eq:folding-condition}
    2\pi \antspacing^{-1} \geq \spatialband(\testedloc, \sourceloc).
\end{equation}
Reciprocally, given $\sourceloc$, aliasing can appear in the \gls{af} for the values of $\testedloc$ that do not satisfy \eqref{eq:folding-condition}.

\subsection{Chirp-based analysis}
The chirp-based approach we present here provides a tractable methodology to approximate $\spatialband(\sourceloc, \testedloc)$, enabling us to develop from \eqref{eq:folding-condition} the aliasing properties affecting the \gls{af} associated with a discrete antenna array.

Defining the local phase and amplitude of $\sigprod$ respectively as
\begin{align}
    \label{eq:chirp-phase}
    \sigprodphase(\paramvar) &:= \sourcek\big(\|\antloc(\paramvar) - \sourceloc\| - \|\antloc(\paramvar) - \testedloc\|\big)\\
    \sigprodamp(\paramvar) &:= \frac{\left\| \dot \antloc(\paramvar)  \right\|}{\left\| \antloc(\paramvar) - \sourceloc \right\|\left\| \antloc(\paramvar) - \testedloc \right\|},
    \label{eq:chirp-amplitude}
\end{align}
we can write
\begin{equation}
    \label{eq:sigprod-2}
    \sigprod(\paramvar \seppar \testedloc, \sourceloc)
    =
    \sigprodamp(\paramvar) \exp\big(-j \sigprodphase(\paramvar)\big).
\end{equation}
We interpret the non-linear phase content $e^{-j \sigprodphase(\paramvar)}$ as a spatial ``chirp", by reference to time chirp signals that are commonly used, for example, in radar systems~\cite{chassande_stationary_1998}. 
Formally, a \emph{spatial} (resp. \emph{time}) chirp is defined by a varying \emph{local wave number} (resp. \emph{instantaneous frequency}).
In our context, this local wave number is given by the derivative $\dot\sigprodphase(\paramvar)$ of $\sigprodphase(\paramvar)$.
As already suggested in the literature~\cite{chassande_stationary_1998}, it is relevant to assume that the set $\{\dot\sigprodphase(\paramvar) : \paramvar \in \paramdomain\}$ observed by the array is a good indication of the ``most active" spatial frequencies (wave numbers) in the chirp's spectrum.

Assuming a high-frequency source signal (\ie a large $\sourcek$), the chirp factor $e^{-j \sigprodphase(\paramvar)}$ oscillates much faster than $\sigprodamp(\paramvar)$, allowing us to neglect the latter's convolutive effect that widens the support of $\spectrumprod$ compared to the pure chirp's spectrum.
In that case, leveraging the compressive spectral modeling of the chirp provides a good approximation of the band limit as
\begin{equation}
    \label{eq:chirp-bandlimit}
    \spatialband(\testedloc, \sourceloc) \simeq \max\{|\dot \sigprodphase(\paramvar)| : \paramvar \in \paramdomain\}.
\end{equation}
Studying cases where the effect of $\sigprodamp(\paramvar)$ must be taken into account in $\spatialband(\testedloc, \sourceloc)$ is postponed to future work.

Our methodology can be summarized as follows: 
\begin{enumerate}
    \item Given a specific array geometry $\antloc(\paramvar)$, we approximate the spatial band limit $\spatialband(\testedloc, \sourceloc)$ with \eqref{eq:chirp-bandlimit}.
    \item We then exploit \eqref{eq:folding-condition} to identify the locus of $\testedloc$ values outside which aliasing (degrading the \gls{af}) appears.
\end{enumerate}

We can check how this generalized methodology trivially applies to the classic 1D \gls{ula} in the \gls{ff} regime. 
Indeed, with $\sourceangle = \angle(\sourceloc)$ and $\mtest\sourceangle = \angle(\testedloc)$, the Fraunhofer approximation gives a ``chirp" which is degenerated to a single frequency content $\dot\sigprodphase(\paramvar) = \sourcek (\cos(\sourceangle) - \cos(\mtest\sourceangle))$, straightforwardly providing $\spatialband(\testedloc, \sourceloc)$.
Therefore, the condition \eqref{eq:folding-condition} simplifies to $2\pi\antspacing^{-1} \geq \sourcek |\cos(\sourceangle) - \cos(\mtest\sourceangle)|$ which can only be satisfied for all angles if the standard condition $\antspacing \leq \sourcewavelen/2$ is met.

In the next two sections, we apply this methodology to the \gls{ca} and the \gls{ula} topologies in the \gls{swr}.

\section{Circular Array}
\label{sec:ca}
 
Let us first apply our methodology to a circular array. 
It is a canonical configuration, captured by a limited number of parameters and to be used as a seed for more complex situations. 
We use the subscript ``$\ca$" to specify the particularizations of the general quantities defined in the section \ref{sec:chirp-methodology}.

\begin{figure}[tb]
\begin{center}
\includegraphics[width=0.6\linewidth]{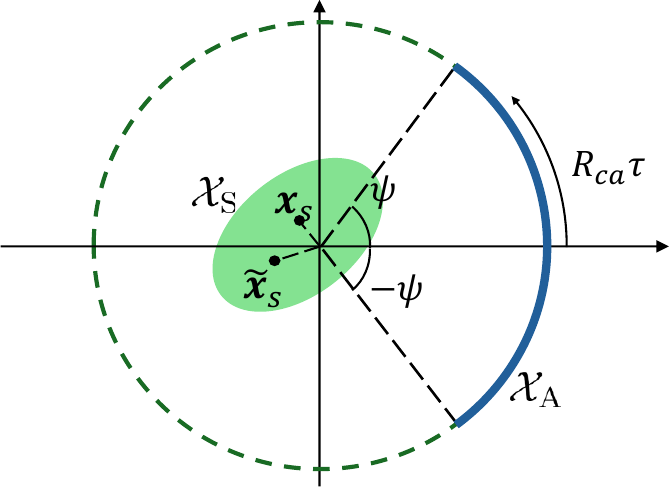}
\end{center}
\caption{Representation of a circular array}
\label{cfscenario}
\end{figure}

Formally, we consider that $\antdomain$ defines an arc on the zero-centered circle of radius $\caradius$ surrounding the domain $\sourcedomain$ of admissible values for the source location $\sourceloc$, as illustrated in Fig.~\ref{cfscenario}.
Without loss of generality, we define this arc as spanning the angular aperture $\paramdomain_\ca := [-\aperture, \aperture]$. 
We define the parametric variable $\paramvar$ to correspond here to the angle such that 
\begin{equation}
    \antloc_\ca(\paramvar) = \caradius [\cos(\paramvar),\, \sin(\paramvar)]^\top.
\end{equation}

\subsection{Chirp-based analysis of the aliasing in a CA}

\begin{figure}[!thb]
    \centering
    \includegraphics[width=\linewidth]{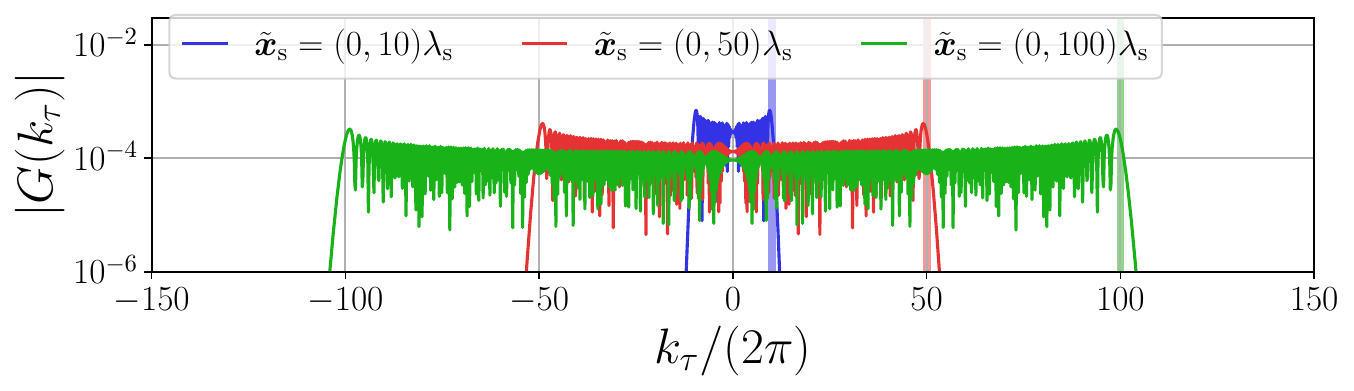}
    \includegraphics[width=\linewidth]{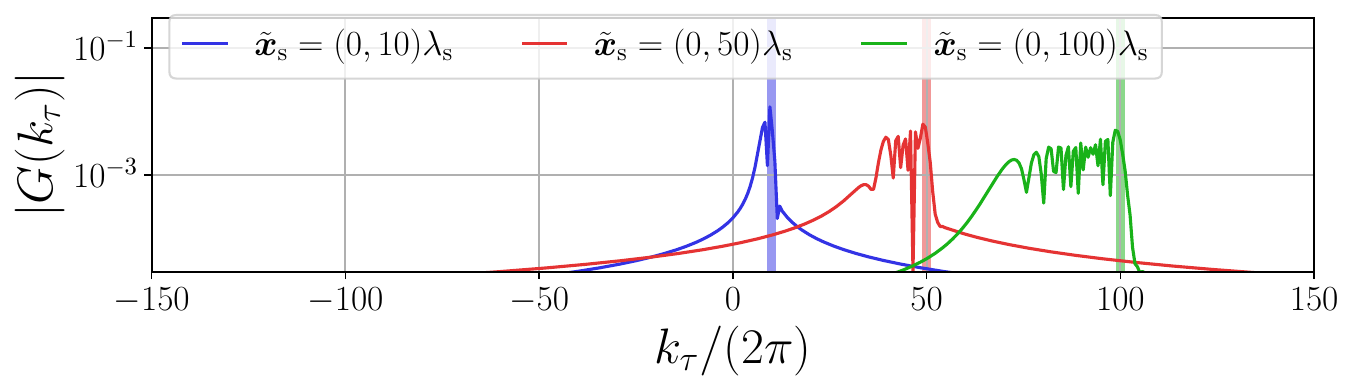}
    \includegraphics[width=\linewidth]{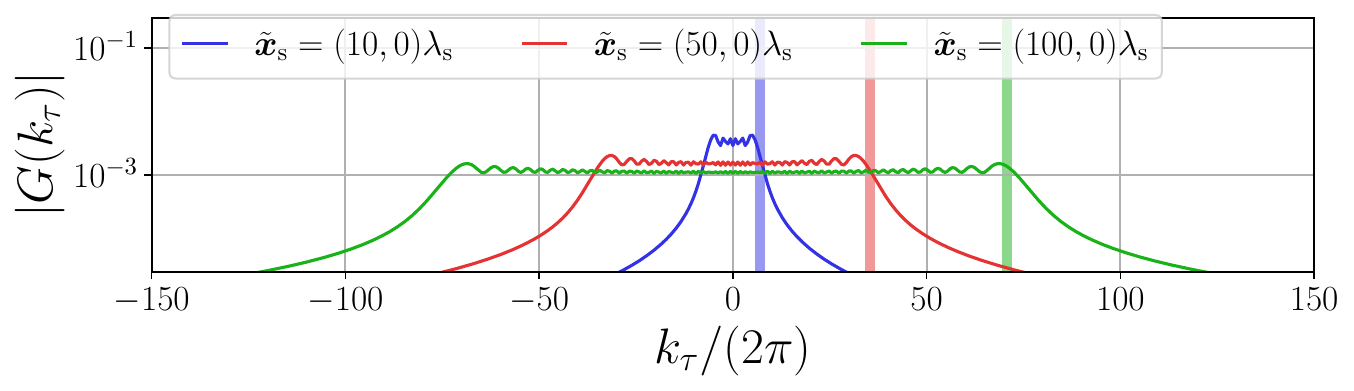}
    \caption{Spectrum $\spectrumprod_\ca(\paramfreq)$ computed for \glspl{ca} with $\caradius=1000\sourcewavelen$, and with $\aperture = \pi$ (top) or $\aperture = 0.25\pi$ (mid and bottom), given $\sourceloc = \bs 0$ fixed.
    The vertical colored bands are the band limits provided by \eqref{eq:spatial-band-ca}.
    In the mid and bottom figures, $\diffangle$ respectively provides $\visualaperture(\diffangle) = 1$ and $\visualaperture(\diffangle) = 0.7$.
    }
    \label{fig:uca-spectrum}
\end{figure}

\begin{figure*}[!tb]
    \centering
    \includegraphics[width=0.95\linewidth]{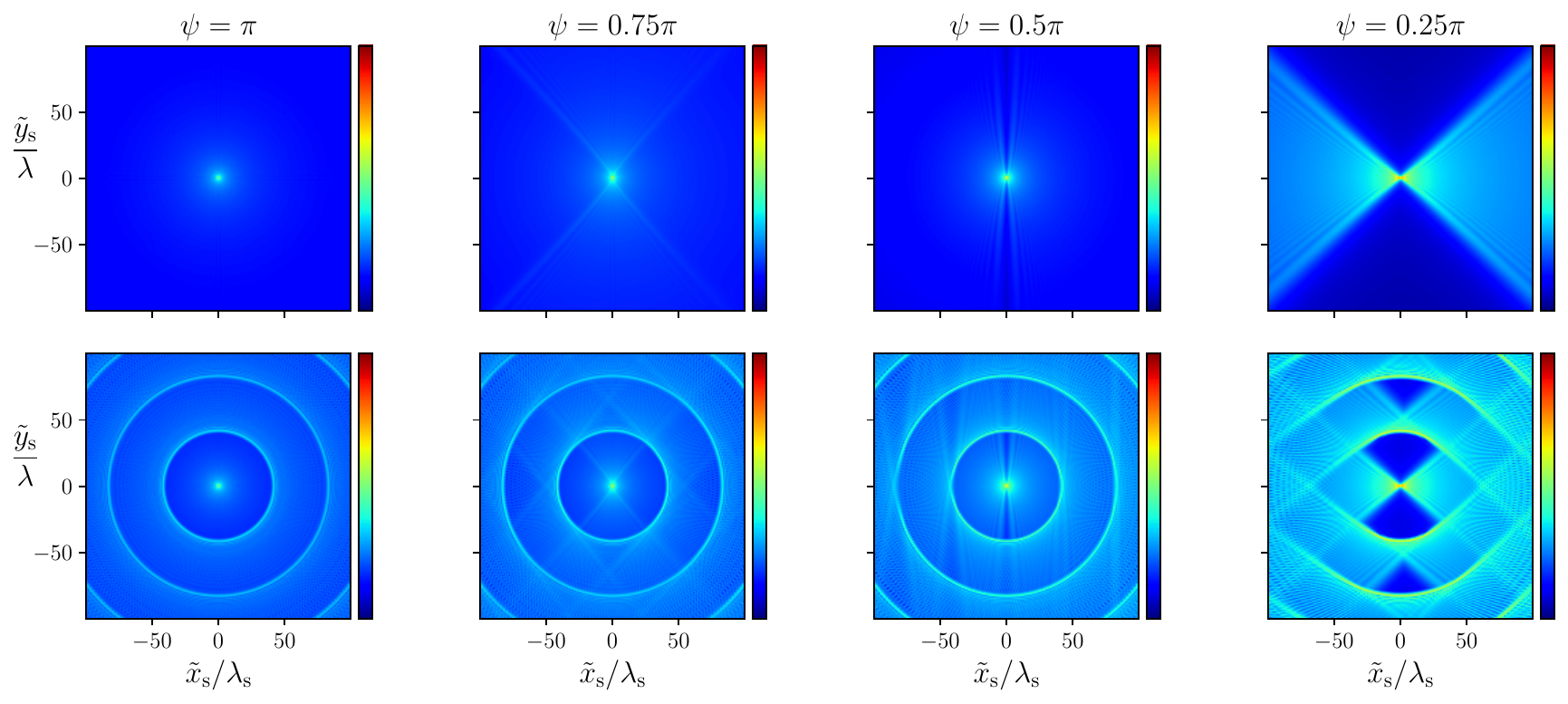}
    \caption{Representation of $|\ambifun(\testedloc, \sourceloc=\mathbf 0)|^\frac1{2}$ (with the exponent $1/2$ only for visibility), for a \gls{ca} of radius $\caradius = 1000 \sourcewavelen$ with different apertures $\aperture$. 
    The top (resp. bottom) figures are obtained with uniform angular spacing $\antspacing_\ca = 2\pi/4096$ (resp. $\antspacing_\ca = 2\pi/256$) between the antennas on the (arc of) circle.
    }
    \label{fig:CA-AF}
\end{figure*}

To develop further the expression of the phase \eqref{eq:chirp-phase}, we focus on scenarios such that $\caradius \gg \max\{ \|\sourceloc\| : \sourceloc \in \sourcedomain \}$.
This assumption has no impact on the \gls{swr} context as we still have to model, in $\sig(\antloc(\paramvar)\seppar\sourceloc)$, the spherical front transmitted by the source. 
However, in the reciprocal point of view, it implies that each specific \gls{aae} is seen in the \gls{ff} of the domain $\sourcedomain$. 
Therefore, for all $\antloc\in\antdomain$ and for all $\testedloc, \sourceloc\in\sourcedomain$, the conventional cosine approximation holds, enabling us to write~\cite{vandendorpe_positioning_2025} 
\begin{equation}
    \label{eq:phase-approx-ca}
    \sigprodphase_\ca(\paramvar) 
    \simeq \sourcek\diffradius \cos(\paramvar -\diffangle),
\end{equation}
where we define
$
    \diffradius = \|\sourceloc - \testedloc\| \text{ and }
    \diffangle = \angle (\sourceloc - \testedloc).
$
Moreover, under this condition, the denominator in \eqref{eq:signal-model} can be approximated as $\caradius$ for all $\antloc$ and all $\sourceloc$, leading to a constant value for $\sigprodamp_\ca(\paramvar) \simeq \caradius^{-1}$. 
With this approximation, it comes
\begin{equation}
    \dot\sigprodphase_\ca(\paramvar)
    =
    \sourcek \diffradius \sin(\paramvar-\diffangle),
    \label{eq:spectrumca}
\end{equation}
and thus, applying \eqref{eq:chirp-bandlimit} yields
\begin{align}
    \spatialband_\ca(\testedloc, \sourceloc)
    &= 
    \sourcek \diffradius\: \visualaperture(\diffangle),
    \label{eq:spatial-band-ca}
\end{align}
where $\visualaperture(\diffangle)$ is the \emph{visual aperture} of the array as seen by $\diffangle$, defined as
\begin{equation}
    \visualaperture(\diffangle) := \ts \max_{\paramvar \in[-\aperture, \aperture]} |\sin(\paramvar-\diffangle)|.
\end{equation}
Fig.~\ref{fig:uca-spectrum} shows how expression \eqref{eq:spatial-band-ca} approximating the band limit properly captures the exact spectrum $\spectrumprod_\ca(\paramfreq)$. 

We now consider $N$ antennas uniformly sampling the arc with a spacing $\antspacing_\ca = 2\aperture/N$.
Using \eqref{eq:folding-condition} with the band limit \eqref{eq:spatial-band-ca}, we conclude that aliasing appears in the resulting \gls{af} when evaluated at positions such that $\spatialband_\ca(\testedloc, \sourceloc) > 2\pi\antspacing_\ca^{-1}$, or equivalently
\begin{equation}
    \label{eq:ca-diffradiusmax}
    \diffradius > \diffradiusmax := \frac {\sourcewavelen}{\antspacing_\ca \visualaperture(\diffangle)}.
\end{equation}
Additional aliasing layers will also appear at every multiple of $\diffradiusmax$.

Interestingly, as soon as $\aperture \geq \frac{\pi}{2}$, the visual aperture provides its maximum value $\visualaperture(\diffangle) =1$ for all $\diffangle$, yielding an aliasing front identical in all directions around $\sourceloc$.
In that case, we are therefore expecting a circular aliasing pattern around $\sourceloc$.
When $\aperture < \frac{\pi}{2}$, then $\visualaperture(\diffangle) =1$ only when either $\diffangle + \frac{\pi}{2}$ or $\diffangle - \frac{\pi}{2}$ is in $[-\aperture, \aperture]$.

This can be seen in Fig.~\ref{fig:CA-AF}, which depicts the \emph{exact}, approximation-free, \glspl{af} obtained by fixing $\sourceloc=0$.
The top figures are obtained with $\antspacing_\ca = \pi/2048$ (thus $N=4096$ antennas for the full-aperture) and act as the reference non-aliased \gls{af}, while the bottom ones are obtained with $\antspacing_\ca = \pi/128$, causing the expected circular aliasing at every radius multiple of $\frac{128}{\pi} \sourcewavelen$ following \eqref{eq:ca-diffradiusmax}. 
Stated otherwise, the \gls{af} remains unaffected by the sampling as long as $\diffradius < \frac{128}{\pi} \sourcewavelen$.

When $\aperture = \tfrac{\pi}{4}$, we clearly observe the circular pattern only for angles $\diffangle$ with $\visualaperture(\diffangle) =1$ following our explanation above, and a stretching towards the other angles.

\subsection{Ambiguity function of a CA}
\label{sec:ca-continuous}

Before turning to \glspl{ula}, we further develop the expression of the \gls{af} associated with a \gls{ca}. 
This allows us to strengthen the validation of our approach since, interestingly, approximation \eqref{eq:phase-approx-ca} leads to a closed-form expression for both the continuous and the discrete array \glspl{af}, as detailed in \cite{vandendorpe_positioning_2025}.
With the expressions of $\sigprodamp_\ca(\paramvar)$ and $\sigprodphase_\ca(\paramvar)$ resulting from this approximation, it comes
\begin{align}
    \label{eq:sigprod-ca}
    \sigprod_\ca(\paramvar \seppar \testedloc, \sourceloc) 
    &\simeq
    \caradius^{-1} \exp\big( -j \sourcek \diffradius \cos(\paramvar -\diffangle) \big) \\
    &=
    \caradius^{-1} \sum_{n=-\infty}^{\infty}
    e^{jn(\tfrac{\pi}{2} - \diffangle)} \, \bessel_n(-\sourcek \diffradius ) 
    e^{jn\paramvar},\nonumber
    \label{eq:jacobi-expansion}
\end{align}
where the last equality expands the complex exponential as a sum of Bessel functions using the Jacobi-Anger expansion.
Inserting the above in \eqref{eq:sigprod-ft} and evaluating it in $\paramfreq=0$ provides the approximation $\mapprox\ambifun_\ca(\testedloc, \sourceloc) \simeq \ambifun_\ca(\testedloc, \sourceloc)$.
The integration over $\paramvar$ only applies to the $e^{jn\paramvar}$ factor, leading to
\begin{align}
    \mapprox\ambifun_\ca(\testedloc, \sourceloc) 
    &=
    \frac{2}{\caradius}
    \sum_{n=-\infty}^{\infty}
    e^{jn(\tfrac{\pi}{2} - \diffangle)} \, \bessel_n(-\sourcek \diffradius ) 
    \frac{\sin(n \aperture)}{n}.\nonumber
\end{align}
This expression provides a good approximation of the \glspl{af} shown in the top panels of Fig.~\ref{fig:CA-AF}.

With the full aperture $\aperture=\pi$, we have
\begin{equation}
    \mapprox\ambifun_\ca(\testedloc, \sourceloc)
    =
    \bessel_0(\sourcek \diffradius ).
    \label{eq:ambifun-ca-bessel-0}
\end{equation}
The resolution associated with this \gls{af}, measured as the argument for its first zero crossing, is $R_{s\tilde{s}}\simeq 2.4 \sourcewavelen/2 \pi$, \ie on the order of the wavelength, as observed in Fig.~\ref{fig:CA-AF}-(top-left).

Considering now the discrete \gls{ca} leads to a summation of $e^{jn\paramvar}$ over $\paramvar_i$, and hence
the approximated \gls{af} becomes 
\begin{align}
    \mapprox\ambifun_{\ca}(\testedloc, \sourceloc)   
    &=
    \sum_{n=-\infty}^{\infty}
    e^{jn(\tfrac{\pi}{2} - \diffangle)} \, \bessel_n(-\sourcek \diffradius )
    \frac{\sin(n \aperture)}{\sin(\tfrac{n}{N} \aperture)}.\nonumber
    \label{ambig3N}
\end{align}
In the full aperture scenario with $\aperture=\pi$, it simplifies to
\begin{equation}
    \mapprox\ambifun_{\ca}(\testedloc, \sourceloc) 
    =
    \sum_{p=-\infty}^{\infty}
    e^{jpN(\tfrac{\pi}{2} - \diffangle)} \, \bessel_{pN}(-\sourcek \diffradius ). 
    \label{ambig3fullN}
\end{equation}
Compared to (\ref{eq:ambifun-ca-bessel-0}), the expression (\ref{ambig3fullN}) exhibits additional terms for $p \neq 0$. 
These are the aliasing terms caused by the folding in $\paramfreq=0$ of the spectrum $\spectrumprod_\ca(\paramfreq \seppar \testedloc, \sourceloc)$.

\section{Uniform linear array the NF regime}
\label{sec:ula}

\begin{figure}[!tb]
    \begin{center}
    \includegraphics[width=0.9\linewidth]{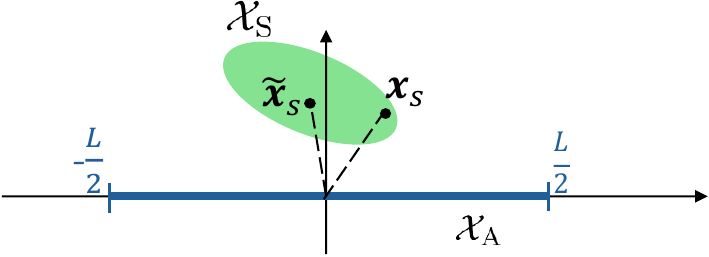}
    \end{center}
    \caption{Representation of a \gls{ula} in the spherical wavefront regime}
    \label{ULAscenario}
    \vspace{-5mm}
\end{figure}

\begin{figure}[!tb]
    \centering
    \includegraphics[width=\linewidth]{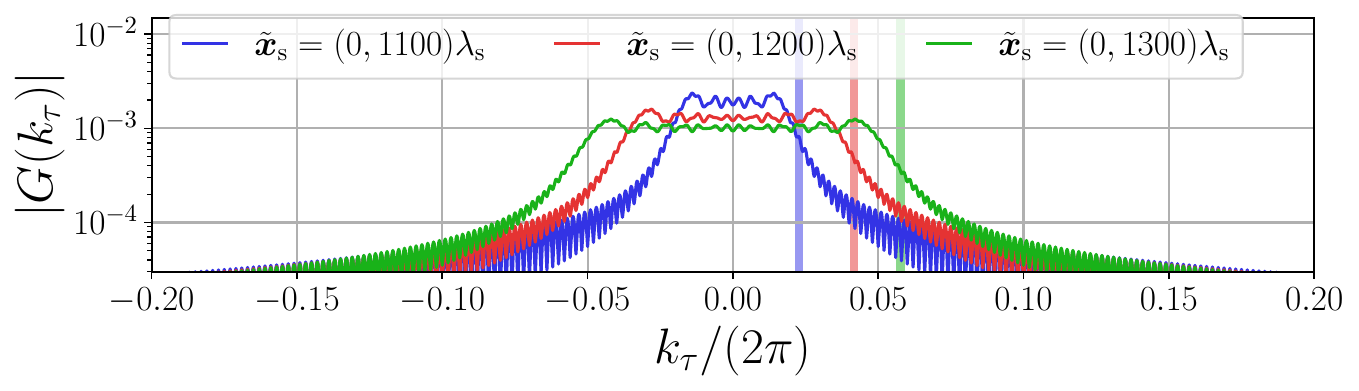}
    \includegraphics[width=\linewidth]{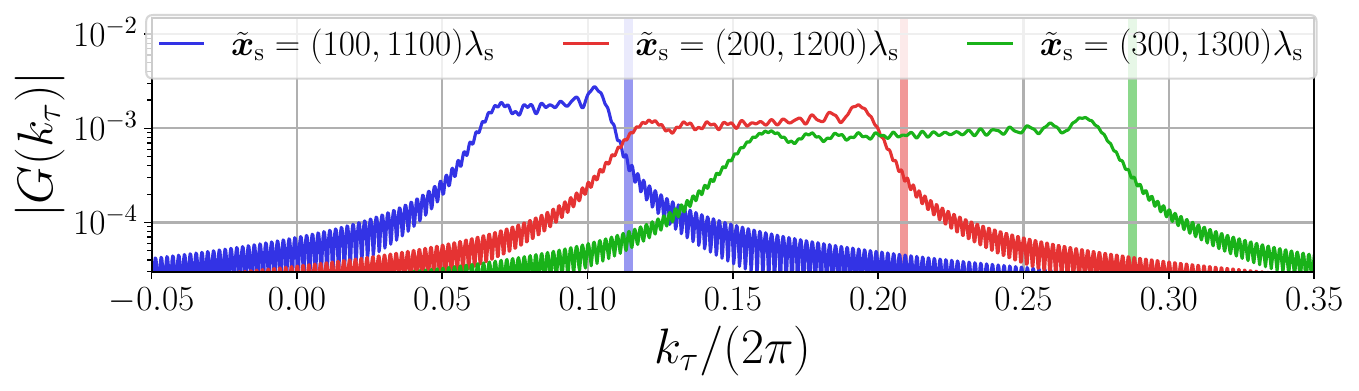}
    \caption{Spectrum $\spectrumprod_\ula(\paramfreq)$ associated with a \gls{ula} with $\lenula = 500\sourcewavelen$ for $\sourceloc = (0, 1000\sourcewavelen)$ and different values of $\testedloc$.
    The band limits estimated by \eqref{rangenf} are represented by the colored vertical bands.
    }
    \label{fig:ula-spectrum}
\end{figure}

Finally, we apply our methodology to a horizontal \gls{ula} of length $\lenula$ centered at the origin, as illustrated in Fig. \ref{ULAscenario}.
It is defined by $\antloc(\paramvar) = [\paramvar, 0]^\top$ for $\paramvar \in \paramdomain_\ula:=[-\lenula/2, \lenula/2]$. 
To get insight to capture the \gls{nf} effects, we use the seminal order-2 (parabolic) Taylor series expansion of the distances
$\|\antloc(\paramvar) - \sourceloc\|$~\cite{liu_near_field_2023, vandendorpe_positioning_2025}, also known as Fresnel approximation. 
With $(\sourceradius, \sourceangle)$ denoting the polar description of the location $\sourceloc$, it comes
 \begin{eqnarray}
\|\antloc(\paramvar) - \sourceloc\| 
 &=&
 [\paramvar^2 + \sourceradius^2 - 2 \paramvar\sourceradius\cos(\theta_s)]^{0.5}
 \nonumber 
 \\
 &\simeq&
 \sourceradius+\paramvar^2/(2\sourceradius) - \paramvar\cos(\theta_s).
 \label{nfdistance}
\end{eqnarray}
Using this approximation, and denoting 
\begin{equation}
    \Delta_\ula:=\frac1{\sourceradius}-\frac1{\tilde{R}_s}
    \text{ and } 
    \Omega_\ula:= \cos(\sourceangle) -\cos(\tilde{\theta}_s),
\end{equation}
the local wave number is given by
\begin{equation}
     \dot\sigprodphase_\ula(\paramvar)
     =
     -\sourcek
     (\paramvar
     \Delta_\ula-
     \Omega_\ula).
     \label{freqnf1}
 \end{equation}
As a result, we obtain from \eqref{eq:def-band-limit} the spatial band limit
\begin{align}
     \spatialband_\ula(\testedloc, \sourceloc)
    &=\sourcek(\tfrac{\lenula}{2}|\Delta_\ula| + |\Omega_\ula|).
     \label{rangenf}
\end{align}
We validate this approximation of the band limit by observing Fig.~\ref{fig:ula-spectrum} which shows the spectrum $\spectrumprod_\ula(\paramfreq)$ for specific values of $\testedloc$, $\sourceloc$ and by comparing it to the expression \eqref{rangenf}.

Now assuming the \gls{ula} is uniformly discretized with $N$ antennas, we apply \eqref{eq:folding-condition} and conclude that the resulting \gls{af} exhibits aliasing for pairs $\sourceloc$, $\testedloc$ that meet
\begin{equation}
     \frac{2\pi N}{\lenula} \leq \sourcek(\tfrac{\lenula}{2}|\Delta_\ula| + |\Omega_\ula|).
     \label{eq:ula-folding}
\end{equation}
This expression highlights the non-space-invariant behavior of the \gls{af} in this \gls{ula} regime.

\begin{figure}[!tb]
    \centering
    \includegraphics[width=\linewidth]{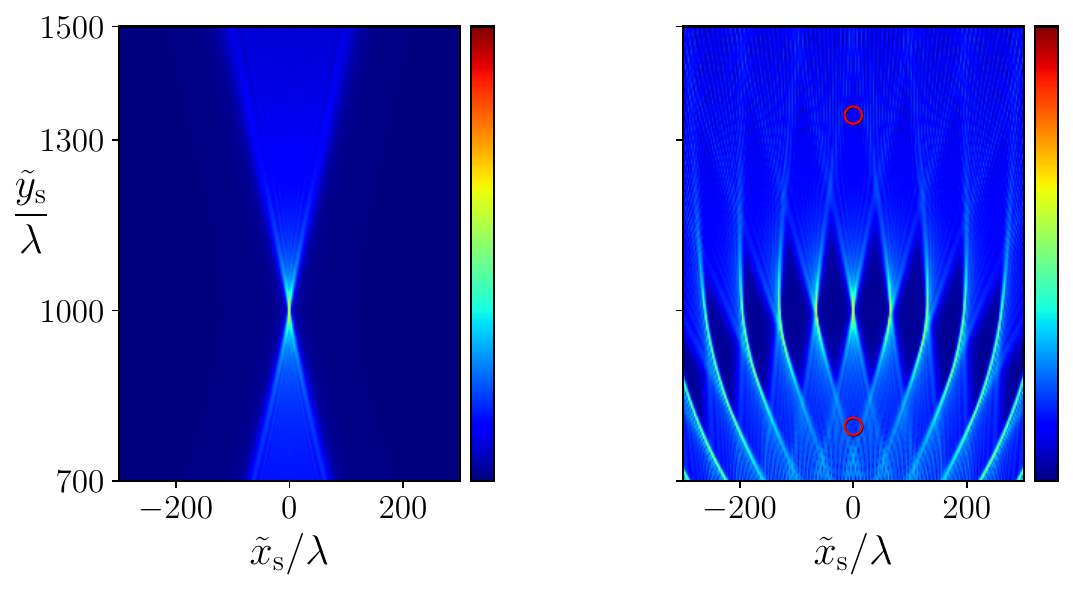}
    \caption{Representation of \glspl{af} obtained for an \gls{ula} with $\lenula = 500\sourcewavelen$ for $\sourceloc = (0, 1000\sourcewavelen)$. We have (left) $N=256$ and (right) $N=32$.}
    \label{fig:ula-af}
\end{figure}

Finally, we show in Fig.~\ref{fig:ula-af} the \glspl{af} associated with the same context as Fig.~\ref{fig:ula-spectrum}.
We interpret the aliasing geometry described by \eqref{eq:ula-folding} as follows. 
In the angular direction ($\Delta_\ula = 0$), the \gls{af}'s peak in $\testedloc = \sourceloc$ is repeated following the \gls{ff}'s conventional cosine rule. 
Radially ($\Omega_\ula = 0$), the \gls{nf} effect appears since, unlike the \gls{ff} regime, these peaks are not degenerated to angular sectors with width driven by the angular resolution. 
Instead, curved cone patterns repeat. 
Using \eqref{eq:ula-folding} for $\Omega_\ula = 0$, we can determine the values of $\mtest\sourceradius$ in which the repeated cones will meet one another and hence radially close the diamond-shaped aliasing locus. 
More precisely, we deduce from \eqref{eq:ula-folding} that the aliasing radially appears for 
\begin{equation}
    \mtest\sourceradius \leq \left(\frac{1}{\sourceradius} + 2\frac{N\sourcewavelen}{\lenula^2}
    \right)^{-1} 
    \text{ and } 
    \mtest\sourceradius \geq \left(\frac{1}{\sourceradius} - 2\frac{N\sourcewavelen}{\lenula^2}
    \right)^{-1}.
    \label{eq:ula-radial-aliasing}
\end{equation}
The scenario of Fig. \ref{fig:ula-af} (right) yields lower and upper limits equal to 796$\sourcewavelen$ and 1344$\sourcewavelen$, highlighted by the red bullets in the figure, which closely match the true aliasing pattern's intersections in the radial direction.

\section{Conclusions}

This paper has been devoted to \glspl{mas} operating in the spherical wavefront regime. 
The steering signal received from a point source along a 1D array has been considered as a spatial chirp described by a set of local wave numbers linked to the array geometry and the source location. 
We leveraged this approach to provide a generalized framework enabling us to understand the structure of the spatial aliasing caused by the array discretization.
We applied it to canonical \glspl{ca} and \glspl{ula}.
For both cases, the dependence of the controllable position's range with respect to the spatial sampling step has been derived, and its validity, demonstrated. 
Future research will be devoted to the impact of the signal bandwidth and of other discrete array geometries on the \gls{af}.

\bibliographystyle{plain}
\bibliography{nourl, references-NF-SL}

\begin{thebibliography}{10}

\bibitem{chassande_stationary_1998}
E.~Chassande-Mottin and P.~Flandrin.
\newblock On the stationary phase approximation of chirp spectra.
\newblock In {\em Proceedings of the {IEEE}-{SP} {International} {Symposium} on {Time}-{Frequency} and {Time}-{Scale} {Analysis} ({Cat}. {No}.{98TH8380})}, pages 117--120, October 1998.

\bibitem{chen_6g_2024}
Hui Chen, Musa~Furkan Keskin, Adham Sakhnini, Nicolò Decarli, Sofie Pollin, Davide Dardari, and Henk Wymeersch.
\newblock {6G} {Localization} and {Sensing} in the {Near} {Field}: {Features}, {Opportunities}, and {Challenges}.
\newblock {\em IEEE Wireless Communications}, 31(4):260--267, 2024.

\bibitem{demir_foundations_2021}
Ozlem~Tugfe Demir, Emil Bjornson, and Luca Sanguinetti.
\newblock Foundations of {User}-{Centric} {Cell}-{Free} {Massive} {MIMO}.
\newblock {\em SIG}, 14(3-4):162--472, January 2021.

\bibitem{jian_fractional_2024}
Mengnan Jian, Anzheng Tang, Yijian Chen, and Yajun Zhao.
\newblock Fractional {Fourier} {Transformation} {Based} {XL}-{MIMO} {Near}-{Field} {Channel} {Analysis}.
\newblock In {\em 2024 {IEEE} 25th {International} {Workshop} on {Signal} {Processing} {Advances} in {Wireless} {Communications} ({SPAWC})}, pages 221--225, September 2024.

\bibitem{kosasih_spatial_2025}
Alva Kosasih, Ozlem~Tuğfe Demir, Nikolaos Kolomvakis, and Emil Bjornson.
\newblock Spatial {Frequencies} and {Degrees} of {Freedom}: {Their} roles in near-field communications.
\newblock {\em IEEE Signal Processing Magazine}, 42(1):33--44, January 2025.

\bibitem{liu_near_field_2023}
Yuanwei Liu, Zhaolin Wang, Jiaqi Xu, Chongjun Ouyang, Xidong Mu, and Robert Schober.
\newblock Near-{Field} {Communications}: {A} {Tutorial} {Review}.
\newblock {\em IEEE Open Journal of the Communications Society}, 4:1999--2049, 2023.

\bibitem{qiu_doa_2018}
Wei Qiu, Wenke Wang, and Wenbin Xiao.
\newblock {DOA} {Estimation} of {Near}-field {Passive} {Sources} with {Acoustic} {Array} {Based} on {Fractional} {Fourier} {Transform}.
\newblock In {\em {OCEANS} 2018 {MTS}/{IEEE} {Charleston}}, pages 1--4, October 2018.

\bibitem{swindlehurst_passive_1988}
A.L. Swindlehurst and T.~Kailath.
\newblock Passive direction-of-arrival and range estimation for near-field sources.
\newblock In {\em Fourth {Annual} {ASSP} {Workshop} on {Spectrum} {Estimation} and {Modeling}}, pages 123--128, August 1988.

\bibitem{vandendorpe_positioning_2025}
Luc Vandendorpe, Laurence Defraigne, Guillaume Thiran, Thomas Pairon, and Christophe Craeye.
\newblock Positioning and transmission in cell-free networks: ambiguity function, and {MRC}/{MRT} array gains.
\newblock In {\em {ICASSP} 2025 - 2025 {IEEE} {International} {Conference} on {Acoustics}, {Speech} and {Signal} {Processing} ({ICASSP})}, pages 1--5, April 2025.

\bibitem{wu_multiple_2023}
Zidong Wu and Linglong Dai.
\newblock Multiple {Access} for {Near}-{Field} {Communications}: {SDMA} or {LDMA}?
\newblock {\em IEEE Journal on Selected Areas in Communications}, 41(6):1918--1935, June 2023.

\bibitem{zhang_6g_2023}
Haiyang Zhang, Nir Shlezinger, Francesco Guidi, Davide Dardari, and Yonina~C. Eldar.
\newblock {6G} {Wireless} {Communications}: {From} {Far}-{Field} {Beam} {Steering} to {Near}-{Field} {Beam} {Focusing}.
\newblock {\em IEEE Communications Magazine}, 61(4):72--77, April 2023.

\end{thebibliography}

\end{document}